\documentclass[runningheads]{llncs}
\usepackage{clpcl}
\usepackage[hidelinks]{hyperref}
\hypersetup{colorlinks=true,linkcolor=blue,citecolor=blue,urlcolor=blue}
\usepackage{orcidlink}
\usepackage{version}

\newcommand{\doi}[1]{{doi:\href{https://doi.org/#1}{\nolinkurl{#1}}}}
\renewcommand{\url}[1]{{\href{#1}{\nolinkurl{#1}}}}

\title{A Classical-Logic View on a Paraconsistent Logic}
\author{C.A. Middelburg\,\orcidlink{0000-0002-8725-0197}}
\institute
 {Informatics Institute, Faculty of Science, University of Amsterdam, \\
  Science Park~900, 1098~XH Amsterdam, the Netherlands \\
  \email{C.A.Middelburg@uva.nl}}

\titlerunning
 {A Classical-Logic View on a Paraconsistent Logic}
\authorrunning
 {C.A. Middelburg}

\begin{document}

\maketitle

\begin{abstract}
This paper is concerned with the paraconsistent first-order logic 
\foLPif, Priest's LPQ enriched with an implication connective 
and a falsity constant.
A sequent-style natural deduction proof system for this logic is 
presented and, for this proof system, both a model-theoretic 
justification and a logical justification by means of an embedding into 
first-order classical logic is given.
The given embedding provides in addition a classical-logic explanation 
of this paraconsistent logic.
As a further matter, its use in decidability issues concerning this 
paraconsistent logic is discussed.
The major properties of \foLPif\ concerning its logical consequence 
relation and its logical equivalence relation are also treated.
The paper emphasizes how closely \foLPif\ is related to classical logic.
\begin{classcode}
Primary 03B53; Secondary 03B10, 03B50, 03B25
\end{classcode}
\end{abstract}

\section{Introduction}
\label{sect-intro}

A set of formulas is contradictory if there exists a formula such that 
both that formula and the negation of that formula can be deduced from 
it.
In classical logic, every formula can be deduced from every 
contradictory set of formulas.
A paraconsistent logic is a logic in which not every formula can be 
deduced from every contradictory set of formulas.

In~\cite{Pri79a}, Priest proposed the paraconsistent propositional logic
LP (Logic of Paradox) and its first-order extension LPQ.
The paraconsistent logic considered in this paper, called \foLPif, is 
LPQ enriched with a falsity constant and an implication connective for 
which the standard deduction theorem holds. 
In this paper, a sequent-style natural deduction proof system for 
\foLPif\ is presented.
In addition to the usual model-theoretic justification of the proof 
system, a logical justification by means of an embedding into 
first-order classical logic is given.
Classical logic is used meta-logically here:
the embedding provides a classical-logic explanation of \foLPif.
The embedding can be used, among other things, to determine for a 
fragment for which validity is known to be decidable in classical logic 
whether validity is decidable in $\foLPif(\Sigma)$ too.
This topic is also discussed.

\foLPif\ is essentially the same as \CLuNs~\cite{BC04a}, 
\foJiii~\cite{DOt85a}, and \foLPc~\cite{Pic18a}.
The proof systems for these logics available in the literature are 
Hilbert systems for the first two logics and a Gentzen-style sequent 
system for the last one.
To fill the gap, a natural deduction proof system for \foLPif\ is given
in this paper. 
An important reason to present a justification of this proof system by 
means of an embedding into classical logic is to draw attention to the 
viewpoint that, although it may be convenient to use a paraconsistent 
logic like \foLPif\ if contradictory sets of formulas have to be dealt 
with, classical logic is the ultima ratio of formal reasoning.

The only difference between \CLuNs\ and \foLPif\ is that the former has 
a bi-implication connective and the latter does not have that 
connective.
However, the bi-implication connective of \CLuNs\ is definable in 
\foLPif.
\foJiii\ and \foLPc\ do not have the falsity constant and the 
implication connective of \foLPif.
Instead, each of \foJiii\ and \foLPc\ has a connective that is foreign 
to classical logic.
However, the constants and connectives of \foLPif\ are definable in 
terms of those of each of these logics and vice versa.
That is why it is said that \foLPif\ is essentially the same as these 
logics.
A plus of \foLPif\ is that it does not have a connective that is foreign 
to classical logic.

Major properties of \foLPif\ concerning its logical consequence relation 
and its logical equivalence relation are also treated in this paper.
The properties in question that concern its logical consequence relation 
are generally considered desirable properties of a reasonable 
paraconsistent first-order logic.
It turns out that 13 classical laws of logical equivalence that also 
hold for the logical equivalence relation of \foLPif\ are sufficient to 
distinguish \foLPif\ completely from the infinitely many three-valued 
paraconsistent first-order logics with the desirable properties 
concerning its logical consequence relation referred to above.

The structure of this paper is as follows.
First, the language of the paraconsistent logic \foLPif\ is 
defined (Section~\ref{sect-LP-iimpl-false}).
Next, a sequent-style natural deduction proof system for \foLPif\ is 
presented (Section~\ref{PROOF-RULES}).
After that, a model-theoretic justification of this proof system is 
given (Section~\ref{INTERPRETATION}).
Then, an embedding of \foLPif\ into classical logic is presented and 
its use in decidability issues concerning \foLPif\ is discussed
(Section~\ref{LPF-TO-L}). 
Following this, the major properties of \foLPif\ concerning its logical 
consequence relation and its logical equivalence relation are presented
(Sections~\ref{sect-properties}).
Finally, some concluding remarks are made (Section~\ref{sect-concl}).

\section{The Language of \foLPif}
\label{sect-LP-iimpl-false}

In this section the language of the paraconsistent logic \foLPif\ is 
described.
First, the assumptions which are made about function and predicate
symbols are given and the notion of a signature is introduced.
Next, the terms and formulas of \foLPif\ are defined for a fixed but 
arbitrary signature. 
Thereafter, notational conventions and abbreviations are presented and 
some remarks about free variables and substitution are made.
In coming sections, the proof system of \foLPif\ and the interpretation 
of the terms and formulas of \foLPif\ are defined for a fixed but 
arbitrary signature.

\subsection{Signatures}
\label{SIGNATURES}

It is assumed that the following has been given:
(a)~a countably infinite set $\SVar$ of \emph{variable symbols},
(b)~for each $n \in \Nat$, a countably infinite set $\SFunc{n}$ of 
\emph{function symbols} of \emph{arity} $n$, and,
(c)~for each $n \in \Nat$, a countably infinite set $\SPred{n}$ of 
\emph{predicate symbols} of \emph{arity} $n$.
It is also assumed that all these sets and the set 
$\set{{=},\Not,\CAnd,\COr,\IImpl,\forall,\exists}$ 
are mutually disjoint.
We write $\SSymb$ for the set 
\mbox{$\SVar \union \Union \set{\SFunc{n} \where n \in \Nat} \union
              \Union \set{\SPred{n} \where n \in \Nat}$}.

Function symbols of arity $0$ are also known as constant symbols and
predicate symbols of arity $0$ are also known as proposition symbols.

A \emph{signature} $\Sigma$ is a non-empty subset of 
$\SSymb \diff \SVar$.
We write $\Func{n}(\Sigma)$ and $\Pred{n}(\Sigma)$, 
where $\Sigma$ is a signature and $n \in \Nat$, 
for the sets $\Sigma \inter \SFunc{n}$ and $\Sigma \inter \SPred{n}$, 
respectively.

The language of \foLPif\ will be defined for a fixed but arbitrary 
signature~$\Sigma$.
This language will be called the language of \foLPif\ over $\Sigma$ or 
shortly the language of $\foLPif(\Sigma)$.
The corresponding proof system and interpretation will be called the
proof system of $\foLPif(\Sigma)$ and the interpretation of 
$\foLPif(\Sigma)$.

\subsection{Terms and formulas}
\label{LANGUAGE}

The language of $\foLPif(\Sigma)$ contains terms and formulas.
They are constructed according to the formation rules given below.

The set of all terms of $\foLPif(\Sigma)$, written $\STerm{\Sigma}$,
is inductively defined by the following formation rules:
\begin{enumerate}
\item
if $x \in \SVar$, then $x \in \STerm{\Sigma}$;
\item
if $c \in \Func{0}(\Sigma)$, then $c \in \STerm{\Sigma}$;
\item
if $f \in \Func{n+1}(\Sigma)$ and 
$t_1,\ldots,t_{n+1} \in \STerm{\Sigma}$, then \newline \phantom{if}
$f(t\sb1,\ldots,t_{n+1}) \in \STerm{\Sigma}$.
\end{enumerate}

The set of all formulas of $\foLPif(\Sigma)$, written $\SForm{\Sigma}$, 
is inductively defined by the following formation rules:
\begin{enumerate}
\item
if $p \in \Pred{0}(\Sigma)$, then 
$p \in \SForm{\Sigma}$;
\item
if $P \in \Pred{n+1}(\Sigma)$ and 
$t_1,\ldots,t_{n+1} \in \STerm{\Sigma}$, then \newline \phantom{if}
$P(t_1,\ldots,t_{n+1}) \in \SForm{\Sigma}$;
\item
if $t_1,t_2 \in \STerm{\Sigma}$, then $t_1 = t_2 \in \SForm{\Sigma}$;
\item
$\False \in \SForm{\Sigma}$;
\item
if $A \in \SForm{\Sigma}$, then $\Not A \in \SForm{\Sigma}$;
\item
if $A_1,A_2 \in \SForm{\Sigma}$, then 
$A_1 \CAnd A_2,\, A_1 \COr A_2,\, A_1 \IImpl A_2 \in \SForm{\Sigma}$;
\item
if $x \in \SVar$ and $A \in \SForm{\Sigma}$, then 
$\CForall{x}{A},\, \CExists{x}{A} \in \SForm{\Sigma}$.
\end{enumerate}
The propositional fragment of $\SForm{\Sigma}$, written $\SProp{\Sigma}$,
is the subset of $\SForm{\Sigma}$ inductively defined by the formation 
rules $1$, $4$, $5$, and $6$.
The set of all atomic formulas of $\SForm{\Sigma}$, written 
$\SAtom{\Sigma}$, is the subset of $\SForm{\Sigma}$ inductively defined 
by the formation rules $1$, $2$, and $3$.

For the connectives $\Not$, $\CAnd$, $\COr$, and $\IImpl$ and the 
quantifiers $\forall$ and $\exists$, the classical truth-conditions and 
falsehood-conditions are retained.
Except for implications, a formula is classified as both-true-and-false 
exactly when it cannot be classified as true or false by these 
conditions.

\subsection{Notational conventions and abbreviations}

In the sequel, some notational conventions and abbreviations will be 
used. 

The following will sometimes be used without mentioning (with or without 
subscripts):
$x$~as a syntactic variable ranging over all variable symbols from 
$\SVar$,
$t$~as a syntactic variable ranging over all terms from 
$\STerm{\Sigma}$, 
$A$ as a syntactic variable ranging over all formulas from 
$\SForm{\Sigma}$, and
$\Gamma$ as a syntactic variable ranging over all finite sets of 
formulas from $\SForm{\Sigma}$.

The string representation of terms and formulas suggested by the 
formation rules given above can lead to syntactic ambiguities. 
Parentheses are used to avoid  such ambiguities.
The need to use parentheses is reduced by ranking the precedence of the 
logical connectives $\Not$, $\CAnd$, $\COr$, $\IImpl$.
The enumeration presents this order from the highest precedence to the
lowest precedence.
Moreover, the scope of the quantifiers extends as far as possible to
the right and $\Forall{x_1}{\cdots \Forall{x_n}{A}}$ is usually written 
as $\Forall{x_1,\ldots,x_n}{A}$.

Non-equality, truth, and bi-implication are defined as abbreviations:
$t_1 \neq t_2$ stands for $\Not (t_1 = t_2)$,
$\True$ stands for $\Not \False$, 
$A_1 \BIImpl A_2$ stands for $(A_1 \IImpl A_2) \CAnd (A_2 \IImpl A_1)$.

\subsection{Free variables and substitution}

Free variables of a term or formula and substitution for variables in a 
term or formula are defined in the usual way.
We write $\ifree(e)$, where $e$ is a term from $\STerm{\Sigma}$ or a 
formula from $\SForm{\Sigma}$, for the set of \emph{free variables} of 
$e$.
We write $\ifree(\Gamma)$, where $\Gamma$ is a finite set of formulas 
from $\SForm{\Sigma}$, for $\Union \set{\ifree(A) \where A \in \Gamma}$.

Let $x$ be a variable symbol from $\SVar$, $t$ be a term from 
$\STerm{\Sigma}$, and $e$ be a term from $\STerm{\Sigma}$ or a formula 
from $\SForm{\Sigma}$.
Then $\subst{x \assign t} e$ is the result of replacing the free 
occurrences of the variable symbol $x$ in $e$ by the term $t$, 
avoiding --- by means of renaming of bound variables --- free variables 
becoming bound in $t$.

\section{Proof System of $\foLPif(\Sigma)$}
\label{PROOF-RULES}

The proof system of $\foLPif(\Sigma)$ is formulated as a sequent-style
natural deduction proof system.
This means that the inference rules have sequents as premises and 
conclusions.
First, the notion of a sequent is introduced.
Next, the inference rules of the proof system of $\foLPif(\Sigma)$ are 
presented.
Then, the notion of a derivation of a sequent from a set of 
sequents and the notion of a proof of a sequent are introduced.
An extension of the proof system of $\foLPif(\Sigma)$ which can serve as 
a proof system for first-order classical logic is also described.

\subsection{Sequents}

In $\foLPif(\Sigma)$, a \emph{sequent} is an expression of the form 
$\Gamma \pEnt A$, where $\Gamma$ is a finite set of formulas from 
$\SForm{\Sigma}$ and $A$ is a formula from $\SForm{\Sigma}$.
We write ${} \pEnt A$ instead of $\emptyset \pEnt A$. 
Moreover, we write $\Gamma,\Gamma'$ for $\Gamma \union \Gamma'$ and 
$A$ for $\set{A}$ on the left-hand side of a sequent.

The intended meaning of the sequent $\Gamma \pEnt A$ is that the
formula $A$ is a logical consequence of the formulas $\Gamma$.
There are several sensible notions of logical consequence in the case 
where formulas can be classified as both-true-and-false.
The notion underlying \foLPif\ is precisely defined in 
Section~\ref{INTERPRETATION}.
It corresponds to the intuitive idea that one can draw conclusions that
are not false from premises that are not false.
Sequents are proved by (natural deduction) proofs obtained by using the 
rules of inference given below.

\subsection{Rules of inference}
\label{subsect-proof-system}

\sloppy
The sequent-style natural deduction proof system of $\foLPif(\Sigma)$ 
consists of the inference rules given in Table~\ref{table-proof-system}.
\begin{table}[!t]
\caption{Sequent-style natural deduction proof system of $\foLPif(\Sigma)$}
\label{table-proof-system}
\vspace*{-3ex} \par \mbox{} 
\renewcommand{\arraystretch}{1.25} 
\centering
\begin{tabular}[t]{c}
\hline
\mbox{} \\[-2.5ex]
\begin{small}
\begin{tabular}{@{}l@{}} 
\InfRule{I}
 {{}}
 {\Gamma, A \pEnt A}
\\[3ex] 
\InfRule{$\True$-I}
 {{}}
 {\Gamma \pEnt \Not \False}
\\[3ex] 
\InfRule{$\CAnd$-I}
 {\Gamma \pEnt A\sb1 \quad
  \Gamma \pEnt A\sb2}
 {\Gamma \pEnt A\sb1 \CAnd A\sb2}
\\[3ex] 
\InfRuleC{$\COr$-I}
 {\Gamma \pEnt A\sb{i}}
 {\Gamma \pEnt A\sb1 \COr A\sb2}
 { for $i =1,2$}
\\[3ex]
\InfRule{$\IImpl$-I}
 {\Gamma, A\sb1 \pEnt A\sb2}
 {\Gamma \pEnt A\sb1 \IImpl A\sb2}
\\[3ex]
\InfRuleC{$\forall$-I}
 {\Gamma \pEnt A}
 {\Gamma \pEnt \CForall{x}{A}}
 {\dag}
\\[3ex]
\InfRule{$\exists$-I}
 {\Gamma \pEnt \subst{x \assign t}A}
 {\Gamma \pEnt \CExists{x}{A}}
\\[3ex]
\InfRule{$=$-I}
 {{}}
 {\Gamma \pEnt t = t}
\\[3ex]
\InfRuleD{$\Not$-M}
 {\Gamma \pEnt \Not \Not A}
 {\Gamma \pEnt A}
\\[3ex]
\InfRuleD{$\COr$-M}
 {\Gamma \pEnt \Not (A\sb1 \COr A\sb2)}
 {\Gamma \pEnt \Not A\sb1 \CAnd \Not A\sb2}
\\[3ex]
\InfRuleD{$\forall$-M}
 {\Gamma \pEnt \Not \CForall{x}{A}}
 {\Gamma \pEnt \CExists{x}{\Not A}}
\\[3ex]
\end{tabular}
\qquad
\begin{tabular}{@{}l@{}}
\InfRule{EM}
 {{}}
 {\Gamma \pEnt A \COr \Not A}
\\[3ex] 
\InfRule{$\False$-E}
 {\Gamma \pEnt \False}
 {\Gamma \pEnt A}
\\[3ex] 
\InfRuleC{$\CAnd$-E}
 {\Gamma \pEnt A\sb1 \CAnd A\sb2}
 {\Gamma \pEnt A\sb{i}}
 {for $i =1,2$}
\\[3ex]
\InfRule{$\COr$-E}
 {\Gamma \pEnt A\sb1 \COr A\sb2 \;\;\;
  \Gamma, A\sb1 \pEnt A\sb3 \;\;\;
  \Gamma, A\sb2 \pEnt A\sb3}
  {\Gamma \pEnt A\sb3}
\\[3ex]
\InfRule{$\IImpl$-E}
 {\Gamma \pEnt A\sb1 \IImpl A\sb2 \quad
  \Gamma \pEnt A\sb1}
 {\Gamma \pEnt A\sb2}
\\[3ex]
\InfRule{$\forall$-E}
 {\Gamma \pEnt \CForall{x}{A}}
 {\Gamma \pEnt \subst{x \assign t}A}
\\[3ex]
\InfRuleC{$\exists$-E}
 {\Gamma \pEnt \CExists{x}{A\sb1} \quad
  \Gamma, A\sb1 \pEnt A\sb2}
 {\Gamma \pEnt A\sb2}
 {\ddag}
\\[3ex]
\InfRule{$=$-E}
 {\Gamma \pEnt t\sb1 = t\sb2 \quad
  \Gamma \pEnt \subst{x \assign t\sb1}A}
 {\Gamma \pEnt \subst{x \assign t\sb2}A}
\\[3ex]
\InfRuleD{$\CAnd$-M}
 {\Gamma \pEnt \Not (A\sb1 \CAnd A\sb2)}
 {\Gamma \pEnt \Not A\sb1 \COr \Not A\sb2}
\\[3ex]
\InfRuleD{$\IImpl$-M}
 {\Gamma \pEnt \Not (A\sb1 \IImpl A\sb2)}
 {\Gamma \pEnt A\sb1 \CAnd \Not A\sb2}
\\[3ex]
\InfRuleD{$\exists$-M}
 {\Gamma \pEnt \Not \CExists{x}{A}}
 {\Gamma \pEnt \CForall{x}{\Not A}}
\\[3ex]
\end{tabular}
\end{small}
\\
$\dag$ restriction on rule $\forall$-I: 
$x \notin \ifree(\Gamma)$; 
\phantom{${} \union \set{A\sb2}$}
\\ 
$\ddag$ restriction on rule $\exists$-E: 
$x \notin \ifree(\Gamma \union \set{A\sb2})$.
\vspace*{1.25ex} \\
\hline
\end{tabular}
\vspace*{-1.5ex}
\end{table}
In this table, 
$x$ is a syntactic variable ranging over all variable symbols from $\SVar$,
$t_1$, $t_2$, and $t$ are syntactic variables ranging over all terms 
from $\STerm{\Sigma}$, and
$A_1$, $A_2$, $A_3$, and $A$ are syntactic variables ranging over all 
formulas from $\SForm{\Sigma}$. 
Double lines indicate a two-way inference rule.

\subsection{Derivations and proofs}

In $\foLPif(\Sigma)$, a \emph{derivation} of a sequent $\Gamma \pEnt A$ 
from a finite set of sequents $\mathcal{H}$ is a finite sequence 
$\seq{s_1,\ldots,s_n}$ of sequents such that $s_n$ equals 
$\Gamma \pEnt A$ and, for each $i \in \set{1,\ldots,n}$, one of the 
following conditions holds:
\begin{itemize}
\item
$s_i \in \mathcal{H}$;
\item
$s_i$ is the conclusion of an instance of some inference rule from the 
proof system of $\foLPif(\Sigma)$ whose premises are among 
$s_1,\ldots,s_{i-1}$.
\end{itemize}
A \emph{proof} of a sequent $\Gamma \pEnt A$ is a derivation of 
$\Gamma \pEnt A$ from the empty set of sequents.
A sequent $\Gamma \pEnt A$ is said to be \emph{provable} if there exists 
a proof of $\Gamma \pEnt A$.

An inference rule that does not belong to the inference rules of some 
proof system is called a \emph{derived inference rule} if there exists 
a derivation of the conclusion from the premises, using the inference
rules of that proof system, for each instance of the rule.

The difference between \CLuNs\ and \foLPif\ is that bi-implication is a
logical connective in \CLuNs\ and must be defined as an abbreviation in
\foLPif.
In~\cite{BC04a}, a proof system of \CLuNs\ is presented which is 
formulated as a Hilbert system.
Removing the axiom schemas A$\BIImpl$1, A$\BIImpl$2, and A$\BIImpl$3 
from this proof system and taking formulas of the form $A_1 \BIImpl A_2$ 
in this proof system as abbreviations yields a proof system of \foLPif\ 
formulated as a Hilbert system.
Henceforth, this proof system will be referred to as the H proof system 
of \foLPif\ and the proof system presented in 
Section~\ref{subsect-proof-system} will be referred to as the ND proof 
system  of \foLPif.

\subsection{A proof system of $\FOCL(\Sigma)$}
\label{subsect-FOCL}

The name \FOCL\ is used to denote a version of classical logic that has 
the same logical constants, connectives, and quantifiers as \foLPif.

In \FOCL, the same assumptions about symbols are made as in \foLPif\ and
the notion of a signature is defined as in \foLPif.
The languages of $\FOCL(\Sigma)$ and $\foLPif(\Sigma)$ are the same.
A natural deduction proof system of $\FOCL(\Sigma)$ can be obtained by 
adding the following inference rule to the ND proof system of 
$\foLPif(\Sigma)$:
\\[1.5ex]
\hspace*{1.5em}
\begin{tabular}{@{}c@{}} 
\InfRule{C}
 {\Gamma \pEnt A\sb1 \quad \Gamma \pEnt \Not A\sb1}
 {\Gamma \pEnt A\sb2}\,\,.
\end{tabular}
\\[1.5ex]
 This proof system is known to be sound and complete.%
\footnote
{If we replace the inference rule EM by the inference rule C in the ND
 proof system of $\foLPif(\Sigma)$, then we obtain a sound and complete
 proof system of the paracomplete analogue of \foLPif.
 The propositional fragment of that logic (\Klif) is studied
 in~\cite{Mid17a}.}
There exist better known alternatives to it, but this proof system is 
arguably the most appropriate one in this paper. 

In Section~\ref{LPF-TO-L}, the sequents of $\foLPif(\Sigma)$ will be 
translated to sequents of $\FOCL(\Sigma')$ ($\Sigma'$ is a particular 
signature related to $\Sigma$).
The translation concerned has the property that what can be derived 
remains the same after translation.
This implies that the inference rules of the proof system of 
$\foLPif(\Sigma)$ become derived inference rules of the above-mentioned 
proof system of $\FOCL(\Sigma')$ after translation.
Thus, the translation provides a logical justification for the inference 
rules of $\foLPif(\Sigma)$.
A model-theoretic justification is afforded by the interpretation given 
in Section~\ref{INTERPRETATION}.

\section{Interpretation of Terms and Formulas of $\foLPif(\Sigma)$}
\label{INTERPRETATION}

The proof system of \foLPif\ is based on the interpretation of the terms 
and formulas of $\foLPif(\Sigma)$ presented below: the inference rules 
preserve validity of sequents under this interpretation.
The interpretation is given relative to a structure and an assignment.
First, the notion of a structure and the notion of an assignment are 
introduced.
Next, the interpretation of the terms and formulas of $\foLPif(\Sigma)$ 
is presented.

\subsection{Structures}

The terms from $\STerm{\Sigma}$ and the formulas from $\SForm{\Sigma}$ 
are interpreted in structures which consist of a non-empty domain of 
individuals and an interpretation of every symbol in the signature $\Sigma$ 
and the equality symbol.
The domain of truth values consists of three values: $\VTrue$ (\emph{true}), 
$\VFalse$ (\emph{false}), and $\VBoth$ (\emph{both true and false}).

A structure $\mathbf{A}$ of $\foLPif(\Sigma)$ consists of:
\begin{itemize}
\item
a set $\mathcal{U}\sp\mathbf{A}$, the \emph{domain} of $\mathbf{A}$, 
such that
$\mathcal{U}\sp\mathbf{A} \neq \emptyset$ and 
$\mathcal{U}\sp\mathbf{A} \inter \set{\VTrue,\VFalse,\VBoth} =
 \emptyset$;
\item
for each $c \in \Func{0}(\Sigma)$, 
\\[1ex] \hspace*{20pt}
an element $c\sp\mathbf{A} \in \mathcal{U}\sp\mathbf{A}$;
\vspace*{1ex}
\item
for each $n \in \Nat$,
for each $f \in \Func{n+1}(\Sigma)$, 
\\[1ex] \hspace*{20pt}
a function 
$f\sp\mathbf{A}: 
 \underbrace{\mathcal{U}\sp\mathbf{A} \X \cdots \X
             \mathcal{U}\sp\mathbf{A}}\sb{n+1\; \mathrm{times}} \to
             \mathcal{U}\sp\mathbf{A}$;
\item
for each $p \in \Pred{0}(\Sigma)$, 
\\[1ex] \hspace*{20pt}
an element $p\sp\mathbf{A} \in \set{\VTrue,\VFalse,\VBoth}$;
\vspace*{1ex}
\item
for each $n \in \Nat$,
for each $P \in \Pred{n+1}(\Sigma)$, 
\\[1ex] \hspace*{20pt}
a function 
$P\sp\mathbf{A}:
 \underbrace{\mathcal{U}\sp\mathbf{A} \X \cdots \X
             \mathcal{U}\sp\mathbf{A}}\sb{n+1\; \mathrm{times}} \to
             \set{\VTrue,\VFalse,\VBoth}$;
\item
a function
$\Meq\sp\mathbf{A}:
 \mathcal{U}\sp\mathbf{A} \X \mathcal{U}\sp\mathbf{A} \to
 \set{\VTrue,\VFalse,\VBoth}$
such that, for all $d_1,d_2 \in \mathcal{U}\sp\mathbf{A}$,
\\ \hspace*{20pt}
$\Meq\sp\mathbf{A}(d_1,d_2) \in \set{\VTrue,\VBoth}$ iff $d_1 = d_2$.
\end{itemize}
Instead of $w\sp\mathbf{A}$ we write $w$ when it is clear from the 
context that the interpretation of symbol $w$ in structure $\mathbf{A}$ 
is meant.

\subsection{Assignments}

An assignment in a structure $\mathbf{A}$ of $\foLPif(\Sigma)$ assigns 
elements from $\mathcal{U}\sp\mathbf{A}$ to the variable symbols from 
$\SVar$.
The interpretation of the terms from $\STerm{\Sigma}$ and the formulas 
from $\SForm{\Sigma}$ in $\mathbf{A}$ is given with respect to an 
assignment $\alpha$ in $\mathbf{A}$.

Let $\mathbf{A}$ be a structure of $\foLPif(\Sigma)$.
Then an \emph{assignment} in $\mathbf{A}$ is a function
$\alpha: \SVar \to \mathcal{U}\sp\mathbf{A}$.
For every assignment $\alpha$ in $\mathbf{A}$, variable symbol 
$x \in \SVar$, and element $d \in \mathcal{U}\sp\mathbf{A}$, we write 
$\alpha(x \to d)$ for the assignment $\alpha'$ in $\mathbf{A}$ such that 
$\alpha'(x) = d$ and $\alpha'(y) = \alpha(y)$ if $y \neq x$.

\subsection{Interpretation}

The interpretation of the terms from $\STerm{\Sigma}$ is given by a 
function mapping term $t$, structure $\mathbf{A}$ and assignment 
$\alpha$ in $\mathbf{A}$ to the element of $\mathcal{U}\sp\mathbf{A}$ 
that is the value of $t$ in $\mathbf{A}$ under assignment $\alpha$.
Similarly, the interpretation of the formulas from $\SForm{\Sigma}$ is 
given by a function mapping formula $A$, structure $\mathbf{A}$ and 
assignment $\alpha$ in $\mathbf{A}$ to the element of 
$\set{\VTrue,\VFalse,\VBoth}$ that is the truth value of $A$ in 
$\mathbf{A}$ under assignment $\alpha$.
We write $\Term{t}{\mathbf{A}}{\alpha}$ and 
$\Term{A}{\mathbf{A}}{\alpha}$ for these interpretations.

The interpretation functions for the terms from $\STerm{\Sigma}$ and 
the formulas from $\SForm{\Sigma}$ are inductively defined in 
Table~\ref{table-interpretation}.
\begin{table}[!t]
\caption{Interpretation of the terms and formulas of $\foLPif(\Sigma)$}
\label{table-interpretation}
\vspace*{-3ex} \par \mbox{} 
\renewcommand{\arraystretch}{1.2}
\centering
\begin{array}[t]{rcl}
\hline
\mbox{} \\[-2.5ex]
\Term{x}{\mathbf{A}}{\alpha} & = &
 \begin{array}[t]{l}
 \alpha(x) \;,
 \end{array}
\\
\Term{c}{\mathbf{A}}{\alpha} & = &
 \begin{array}[t]{l}
 c\sp\mathbf{A} \;,
 \end{array}
\\
\Term{f(t\sb1,\ldots,t\sb{n+1})}{\mathbf{A}}{\alpha} & = &
 \begin{array}[t]{l}
 f\sp\mathbf{A}(\Term{t\sb1}{\mathbf{A}}{\alpha},\ldots,
                \Term{t\sb{n+1}}{\mathbf{A}}{\alpha})
 \end{array}
\\[1.5ex]
\Term{p}{\mathbf{A}}{\alpha} & = &
 \begin{array}[t]{l}
 p\sp\mathbf{A} \;,
 \end{array}
\\
\Term{P(t\sb1,\ldots,t\sb{n+1})}{\mathbf{A}}{\alpha} & = &
 \begin{array}[t]{l}
 P\sp\mathbf{A}(\Term{t\sb1}{\mathbf{A}}{\alpha},\ldots,
                \Term{t\sb{n+1}}{\mathbf{A}}{\alpha}) \;,
 \end{array}
\\
\Term{t\sb1 = t\sb2}{\mathbf{A}}{\alpha} & = &
 \begin{array}[t]{l}
 \Meq\sp\mathbf{A}(\Term{t\sb1}{\mathbf{A}}{\alpha},
                   \Term{t\sb2}{\mathbf{A}}{\alpha}) \;,
\end{array}
\\
\Term{\False}{\mathbf{A}}{\alpha} & = &
 \begin{array}[t]{l}
 \VFalse \;,
 \end{array}
\vspace*{.5ex} \\
\Term{\Not A}{\mathbf{A}}{\alpha} & = &
 \left \{
 \begin{array}{l@{\;\;}l}
 \VTrue  & \mathrm{if}\; \Term{A}{\mathbf{A}}{\alpha} = \VFalse \\
 \VFalse & \mathrm{if}\; \Term{A}{\mathbf{A}}{\alpha} = \VTrue \\
 \VBoth  & \mathrm{otherwise},
 \end{array}
 \right.
\vspace*{.5ex} \\
\Term{A\sb1 \CAnd A\sb2}{\mathbf{A}}{\alpha} & = &
 \left \{
 \begin{array}{l@{\;\;}l}
 \VTrue  & \mathrm{if}\; 
           \Term{A\sb1}{\mathbf{A}}{\alpha} = \VTrue\; \mathrm{and}\;
           \Term{A\sb2}{\mathbf{A}}{\alpha} = \VTrue \\
 \VFalse & \mathrm{if}\;
           \Term{A\sb1}{\mathbf{A}}{\alpha} = \VFalse\; \mathrm{or}\;
           \Term{A\sb2}{\mathbf{A}}{\alpha} = \VFalse \\
 \VBoth  & \mathrm{otherwise},
 \end{array}
 \right.
\vspace*{.5ex} \\
\Term{A\sb1 \COr A\sb2}{\mathbf{A}}{\alpha} & = &
 \left \{
 \begin{array}{l@{\;\;}l}
 \VTrue  & \mathrm{if}\; 
           \Term{A\sb1}{\mathbf{A}}{\alpha} = \VTrue\; \mathrm{or}\;
           \Term{A\sb2}{\mathbf{A}}{\alpha} = \VTrue \\
 \VFalse & \mathrm{if}\;
           \Term{A\sb1}{\mathbf{A}}{\alpha} = \VFalse\; \mathrm{and}\;
           \Term{A\sb2}{\mathbf{A}}{\alpha} = \VFalse \\
 \VBoth  & \mathrm{otherwise},
 \end{array}
 \right.
\vspace*{.5ex} \\
\Term{A\sb1 \IImpl A\sb2}{\mathbf{A}}{\alpha} & = &
 \left \{
 \begin{array}{l@{\;\;}l}
 \VTrue  & \mathrm{if}\; 
           \Term{A\sb1}{\mathbf{A}}{\alpha} = \VFalse\; \mathrm{or}\;
           \Term{A\sb2}{\mathbf{A}}{\alpha} = \VTrue \\
 \VFalse & \mathrm{if}\;
           \Term{A\sb1}{\mathbf{A}}{\alpha} \neq \VFalse\; \mathrm{and}\;
           \Term{A\sb2}{\mathbf{A}}{\alpha} = \VFalse \\
 \VBoth  & \mathrm{otherwise},
 \end{array}
 \right.
\vspace*{.5ex} \\
\Term{\Forall{x}{A}}{\mathbf{A}}{\alpha} & = &
 \left \{
 \begin{array}{l@{\;\;}l}
 \VTrue  & \mathrm{if},\;
           \mathrm{for\; all}\;  d \in \mathcal{U}\sp\mathbf{A},\;
           \Term{A}{\mathbf{A}}{\alpha(x \to d)} = \VTrue \\
 \VFalse & \mathrm{if},\;
           \mathrm{for\; some}\; d \in \mathcal{U}\sp\mathbf{A},\;
           \Term{A}{\mathbf{A}}{\alpha(x \to d)} = \VFalse \\
 \VBoth  & \mathrm{otherwise}.
 \end{array}
 \right.
\vspace*{.5ex} \\
\Term{\Exists{x}{A}}{\mathbf{A}}{\alpha} & = &
 \left \{
 \begin{array}{l@{\;\;}l}
 \VTrue  & \mathrm{if},\;
           \mathrm{for\; some}\;  d \in \mathcal{U}\sp\mathbf{A},\;
           \Term{A}{\mathbf{A}}{\alpha(x \to d)} = \VTrue \\
 \VFalse & \mathrm{if},\;
           \mathrm{for\; all}\; d \in \mathcal{U}\sp\mathbf{A},\;
           \Term{A}{\mathbf{A}}{\alpha(x \to d)} = \VFalse \\
 \VBoth  & \mathrm{otherwise}.
 \end{array}
 \right.
\vspace*{1.25ex} \\
\hline
\end{array}
\vspace*{-1.5ex}
\end{table}
In this table, 
$x$ is a syntactic variable ranging over all variable symbols from 
$\SVar$,
$c$ is a syntactic variable ranging over all function symbols from 
$\Func{0}(\Sigma)$,
$f$ is a syntactic variable ranging over all function symbols from
$\Func{n+1}(\Sigma)$ (where $n$ is understood from the context), 
$p$ is a syntactic variable ranging over all predicate symbols from 
$\Pred{0}(\Sigma)$,
$P$ is a syntactic variable ranging over all  predicate symbols from
$\Pred{n+1}(\Sigma)$ (where $n$ is understood from the context), 
$t_1$, \ldots, $t_{n+1}$ are syntactic variables ranging over all terms 
from $\STerm{\Sigma}$, and 
$A_1$, $A_2$, and $A$ are syntactic variables ranging over all formulas 
from $\SForm{\Sigma}$.

The logical consequence relation of $\foLPif(\Sigma)$ is based on the 
idea that a formula $A$ holds in a structure $\mathbf{A}$ under an 
assignment $\alpha$ in $\mathbf{A}$ if 
$\Term{A}{\mathbf{A}}{\alpha} \in \set{\VTrue,\VBoth}$.

Let $\Gamma$ be a finite set of formulas from $\SForm{\Sigma}$ and
$A$ be a formula from $\SForm{\Sigma}$.
Then $A$ is a \emph{logical consequence} of $\Gamma$,
written $\Gamma \LCon A$, iff
for all structures $\mathbf{A}$ of $\foLPif(\Sigma)$,
for all assignments $\alpha$ in $\mathbf{A}$,
$\Term{A'}{\mathbf{A}}{\alpha} = \VFalse$ for some $A' \in \Gamma$ or
$\Term{A}{\mathbf{A}}{\alpha} \in \set{\VTrue,\VBoth}$.
We write ${} \LCon A$ instead of $\emptyset \LCon A$.

As mentioned before, the difference between \CLuNs\ and \foLPif\ is that 
bi-implication is a logical connective in \CLuNs\ and must be defined as 
an abbreviation in \foLPif.
In~\cite{BC04a}, an interpretation of the formulas of \CLuNs\ is 
presented whose restriction to formulas without occurrences of the 
bi-implication connective is essentially the same as the interpretation
of the formulas of \foLPif\ given above.
The soundness and completeness properties for the Hilbert proof system 
of \CLuNs\ proved in~\cite{BC04a} directly carry over to \foLPif.

\begin{theorem}
\label{theorem-sound-complete}
The ND proof system of $\foLPif(\Sigma)$ presented in 
Section~\ref{subsect-proof-system} is sound and complete, i.e.,
for all finite $\Gamma \subseteq \SForm{\Sigma}$ and 
$A \in \SForm{\Sigma}$, $\Gamma \pEnt A$ is provable in
the ND proof system of $\foLPif(\Sigma)$ iff $\Gamma \LCon A$.
\end{theorem}
\begin{proof}
Because it is known from~\cite{BC04a} that these properties hold for the 
H proof system of \foLPif, it is sufficient to prove that, for each 
finite set $\Gamma$ of formulas from $\SForm{\Sigma}$ and each formula 
$A$ from $\SForm{\Sigma}$, $\Gamma \pEnt A$ is provable in the H system 
of $\foLPif(\Sigma)$ iff $\Gamma \pEnt A$ is provable in the ND system 
of $\foLPif(\Sigma)$.

The only if part is straightforwardly proved by induction on the length 
of the proof of $\Gamma \pEnt A$ in the H system, using that
(a)~for each axiom $A'$ of the H system, $\pEnt A'$ can be proved in the 
ND system and 
(b)~for each inference rule of the H system, a corresponding derived 
inference rule of the ND system can be found.

The if part is straightforwardly proved by induction on the length of 
the proof of $\Gamma \pEnt A$ in the ND system, using that 
(a)~the standard deduction theorem holds for the H system, 
(b)~for each inference rule of the ND system different from I, 
$\IImpl$-E, $\forall$-I, and $\exists$-E, there exists a corresponding 
axiom of the H system, 
(c)~for each of the inference rules $\IImpl$-E, $\forall$-I, and 
$\exists$-E, a corresponding derived inference rule of the H system can
be found, and 
(d)~$\pEnt A \IImpl A$ can be proved in the H system.
\qed
\end{proof}

The notion of logical equivalence is a semantic notion that is relevant 
to any logic.
The logical equivalence relation of \foLPif\ is defined as it is defined 
for classical logic.

Let $A_1$ and $A_2$ be formulas from $\SForm{\Sigma}$. 
Then $A_1$ is \emph{logically equivalent} to $A_2$, written 
$A_1 \LEqv A_2$, iff 
for all structures $\mathbf{A}$ of $\foLPif(\Sigma)$,
for all assignments $\alpha$ in $\mathbf{A}$:
\[\Term{A_1}{\mathbf{A}}{\alpha} = \Term{A_2}{\mathbf{A}}{\alpha}.\]

It should be mentioned that, unlike in classical logic, it does not hold 
in three-valued paraconsistent logics that logical equivalence is the 
same as logical consequence and its inverse.

The notions of  validity and satisfiability are also semantic notions 
that are relevant to any logic.
As in classical logic, these notions are closely related in 
$\foLPif(\Sigma)$.

Let $\Gamma$ be a finite set of formulas from $\SForm{\Sigma}$ and $A$ 
be a formula from $\SForm{\Sigma}$.
Then validity and satisfiability of sequents are defined as follows:
\vspace*{-1.125ex} \par
\[
\renewcommand{\arraystretch}{1.125}
\begin{tabular}[t]{rl}
$\Gamma \pEnt A$ is \emph{valid} &
 iff $\Gamma \LCon A$;
\\
$\Gamma \pEnt A$ is \emph{satisfiable} &
 iff $\Gamma \nLCon A \IImpl \False$.
\end{tabular}
\]
We say that $A$ is \emph{valid} iff $\emptyset \pEnt\! A$ is valid and
that $A$ is \emph{satisfiable} iff $\emptyset \pEnt\! A$ is satisfiable.

The way satisfiability and validity are related in \foLPif\ is very 
similar to the way in which they are related in classical logic.
\begin{proposition}
\label{prop-valid-sat}
For all finite $\Gamma \subseteq \SForm{\Sigma}$ and 
$A \in \SForm{\Sigma}$:
\vspace*{-1.125ex} \par
\[
\renewcommand{\arraystretch}{1.125}
\begin{tabular}[t]{rl}
$\Gamma \pEnt A$ is valid &
 iff $\Gamma \pEnt A \IImpl \False$ is not satisfiable;
\\
$\Gamma \pEnt A$ is satisfiable &
 iff $\Gamma \pEnt A \IImpl \False$ is not valid.
\end{tabular}
\]
\end{proposition}
\begin{proof}
This follows immediately from the definitions of validity and 
satisfiability and the interpretation of formulas of the form 
$A \Impl \False$.
\qed
\end{proof}

The above definition of satisfiability for \foLPif\ is also a suitable 
definition of satisfiability for classical logic.
Replacing $A \IImpl \False$ by $\Not A$ in this definition yields an 
equivalent and more customary definition of satisfiability for classical 
logic and a non-equivalent and unsuitable definition of satisfiability 
for \foLPif.
In \foLPif, the kind of negation provided by $A \IImpl \False$ is 
properly included in the kind of negation provided by $\Not A$ in the 
sense that for all formulas $A$ from $\SForm{\Sigma}$:
\[
\renewcommand{\arraystretch}{1.125}
\begin{tabular}[t]{l}
$\Not A \LEqv \Not A \Or A \IImpl \False$ and 
$\Not A \not\LEqv A \IImpl \False$.
\end{tabular}
\]

The notion of consistency is a semantic notion that is in particular 
relevant to paraconsistent logics.
The consistency property is not definable in classical logic.

Let $A_1$ and $A_2$ be formulas from $\SForm{\Sigma}$.
Then $A$ is \emph{consistent} iff 
for all structures $\mathbf{A}$ of $\foLPif(\Sigma)$,
for all assignments $\alpha$ in $\mathbf{A}$:
\[\Term{A_1}{\mathbf{A}}{\alpha} \neq \VBoth.\]

\section{Embedding of $\foLPif(\Sigma)$ into $\FOCL(\Sigma)$}
\label{LPF-TO-L}

The formulas and sequents of $\foLPif(\Sigma)$ are translated in this 
section to formulas and sequents, respectively, of $\FOCL(\TSigma)$, 
where $\TSigma$ is a signature obtained from the signature $\Sigma$ as 
defined below.
The translation concerned provides a uniform embedding of 
$\foLPif(\Sigma)$ into $\FOCL(\TSigma)$: a sequent is provable in 
$\foLPif(\Sigma)$ iff its translation is provable in 
\smash{$\FOCL(\TSigma)$}.
Thus, the translation provides both a classical-logic explanation of 
$\foLPif(\Sigma)$ and a logical justification of its proof system.
Moreover, it can be useful to determine for a fragment for which 
validity or satisfiability of sequents is decidable in $\FOCL(\TSigma)$ 
whether it is decidable in $\foLPif(\Sigma)$ too.

\subsection{Translation}
\label{subsect-translation}

In the translation, it is assumed that 
${\DXEq} \in \SPred{2} \diff \Pred{2}(\Sigma)$ and 
that an injective function from 
$\Union_{n \in \Nat} \Pred{n}(\Sigma)$ to 
$(\Union_{n \in \Nat} \SPred{n}) \diff 
 (\Union_{n \in \Nat} \Pred{n}(\Sigma) \union \set{\DXEq})$ 
has been given.
We write $\denial{\TP}$, 
where $P \in \Union_{n \in \Nat} \Pred{n}(\Sigma)$, 
for the symbol to which $P$ is mapped by this function, and
we write $\denial{A}$, where $A \in \SAtom{\Sigma}$, 
for $A$ with the symbol 
$P \in \Union_{n \in \Nat} \Pred{n}(\Sigma) \union \set{=}$ 
occurring in $A$ replaced by \smash{$\denial{P}$}.
It is further assumed that, for each $n \in \Nat$, for each 
$P \in \Pred{n}(\Sigma)$,\, $\denial{\TP} \in \SPred{n}$. 

The signature $\TSigma$ is defined by 
\[
\TSigma = 
\Sigma \union 
\Union_{n \in \Nat}
 \set{\denial{\TP} \where P \in \Pred{n}(\Sigma)}) \union 
\set{\DXEq}\;.
\]

The translation of formulas of $\foLPif(\Sigma)$ is given 
by the function
\[
\Embed{\ph}{}{} : \SForm{\Sigma} \to \TForm{\TSigma}
\]
inductively defined in Table~\ref{table-translation}. 
\begin{table}[!t]
\caption{Translation of the formulas of $\foLPif(\Sigma)$}
\label{table-translation}
\vspace*{-3ex} \par \mbox{} \centering
\renewcommand{\arraystretch}{1.275}
\begin{array}[t]{rcl}
\hline
\mbox{} \\[-2.5ex]
\Embed{p}{}{} & = & \Tp\;,
\\
\Embed{P(t\sb1 ,\ldots, t\sb{n+1})}{}{} & = &
  \TP(t\sb1 ,\ldots, t\sb{n+1})\;,
\\
\Embed{t\sb1 = t\sb2}{}{} & = & t\sb1 \XEq t\sb2\;,
\\
\Embed{\False}{}{} & = & \False \;,
\\
\Embed{A\sb1 \CAnd A\sb2}{}{} & = &
  \Embed{A\sb1}{}{} \And \Embed{A\sb2}{}{} \;,
\\
\Embed{A\sb1 \COr A\sb2}{}{} & = &
  \Embed{A\sb1}{}{} \Or \Embed{A\sb2}{}{} \;,
\\
\Embed{A\sb1 \IImpl A\sb2}{}{} & = &
  \Embed{A\sb1}{}{} \IImpl \Embed{A\sb2}{}{} \;,
\\
\Embed{\CForall{x}{A}}{}{} & = &
  \Forall{x}{\Embed{A}{}{}} \;,
\\
\Embed{\CExists{x}{A}}{}{} & = &
  \Exists{x}{\Embed{A}{}{}} \;,
\\[1ex]
\Embed{\Not p}{}{} & = & \Not \Tp \Or \denial{\Tp}\;,
\\
\Embed{\Not P(t\sb1 ,\ldots, t\sb{n+1})}{}{} & = &
  \Not \TP(t\sb1,\ldots,t\sb{n+1}) \Or
  \denial{\TP}(t\sb1 ,\ldots, t\sb{n+1})\;,
\\
\Embed{\Not\; t\sb1 = t\sb2}{}{} & = &
  \Not (t\sb1 = t\sb2) \Or t\sb1 \DXEq t\sb2\;,
\\
\Embed{\Not \False}{}{} & = & \Not \False \;,
\\
\Embed{\Not \Not A}{}{} & = & \Embed{A}{}{} \;,
\\
\Embed{\Not (A\sb1 \CAnd A\sb2)}{}{} & = &
  \Embed{\Not A\sb1 \Or \Not A\sb2}{}{} \;,
\\
\Embed{\Not (A\sb1 \COr A\sb2)}{}{} & = &
  \Embed{\Not A\sb1 \And \Not A\sb2}{}{} \;,
\\
\Embed{\Not (A\sb1 \IImpl A\sb2)}{}{} & = &
  \Embed{A\sb1 \And \Not A\sb2}{}{} \;,
\\
\Embed{\Not\, \CForall{x}{A}}{}{} & = &
  \Embed{\Exists{x}{\Not A}}{}{} \;,
\\
\Embed{\Not\, \CExists{x}{A}}{}{} & = &
  \Embed{\Forall{x}{\Not A}}{}{} \;.
\vspace*{1.25ex} \\
\hline
\end{array}
\vspace*{-1.5ex} \par
\end{table}
In this table, 
$x$ is a syntactic variable ranging over all variable symbols from 
$\SVar$,
$c$ is a syntactic variable ranging over all function symbols from 
$\Func{0}(\Sigma)$,
$f$ is a syntactic variable ranging over all function symbols from
$\Func{n+1}(\Sigma)$ (where $n$ is understood from the context), 
$p$ is a syntactic variable ranging over all predicate symbols from 
$\Pred{0}(\Sigma)$,
$P$ is a syntactic variable ranging over all  predicate symbols from
$\Pred{n+1}(\Sigma)$ (where $n$ is understood from the context), 
$t_1$, \ldots, $t_{n+1}$ are syntactic variables ranging over all terms 
from $\STerm{\Sigma}$, and 
$A_1$, $A_2$, and $A$ are syntactic variables ranging over all formulas 
from $\SForm{\Sigma}$.

The intuition is that $\Embed{A}{}{}$ is a formula of $\FOCL(\TSigma)$ 
stating that the formula $A$ of $\foLPif(\Sigma)$ is either true or both 
true and false in $\foLPif(\Sigma)$.

The translation of sequents of $\foLPif(\Sigma)$ is defined as follows:
\begin{eqnarray*}
\Embed{\Gamma \pEnt A}{}{} & = &
 \set{\Embed{A'}{}{} \where A' \in \Gamma} \pEnt
 \Embed{A}{}{}\;,
\end{eqnarray*}

\subsection{Embedding}

An important property of the translation of sequents of 
$\foLPif(\Sigma)$ to sequents of $\FOCL(\TSigma)$ presented above is 
that what can be proved remains the same after translation.
This means that the translation provides a uniform embedding of 
$\foLPif(\Sigma)$ into $\FOCL(\TSigma)$.
\begin{theorem}
\label{theorem-embedding}
For all finite $\Gamma \subseteq \SForm{\Sigma}$ and 
$A \in \SForm{\Sigma}$:
\[
\renewcommand{\arraystretch}{1.125}
\begin{tabular}[t]{l}
$\Gamma \pEnt A$ is provable in $\foLPif(\Sigma)$ \,iff\,
$\Embed{\Gamma \pEnt A}{}{}$ is provable in 
$\FOCL(\TSigma)$.
\end{tabular}
\]
\end{theorem}
\begin{proof}
The only if part is easily proved by induction on the length of a proof 
of $\Gamma \pEnt A$ and case distinction on the last inference rule 
applied, using that the ND proof system for \FOCL($\TSigma$) 
described in Section~\ref{subsect-FOCL} contains all inference rules of 
$\foLPif(\Sigma)$.

The if part is proved by contrapositive.
Let $\mathbf{A}$ be a structure of $\foLPif(\Sigma)$.
Then $\mathbf{A}$ can be transformed into a structure $\mathbf{A\sp{*}}$ 
of \smash{$\FOCL(\TSigma)$} with the property that 
for all atomic formula $A \in \SAtom{\Sigma}$:
\[
\renewcommand{\arraystretch}{1.25}
\begin{tabular}[t]{l}
$\Term{A}{\mathbf{A\sp{*}}}{\alpha} = \VTrue$ and
$\Term{\denial{A}}{\mathbf{A\sp{*}}}{\alpha} = \VFalse$ \,iff\,
$\Term{A}{\mathbf{A}}{\alpha} = \VTrue$,
\\
$\Term{A}{\mathbf{A\sp{*}}}{\alpha} = \VFalse$ and
$\Term{\denial{A}}{\mathbf{A\sp{*}}}{\alpha} = \VTrue$ \,iff\,
$\Term{A}{\mathbf{A}}{\alpha} = \VFalse$,
\\
$\Term{A}{\mathbf{A\sp{*}}}{\alpha} = \VTrue$ and
$\Term{\denial{A}}{\mathbf{A\sp{*}}}{\alpha} = \VTrue$ \,iff\,
$\Term{A}{\mathbf{A}}{\alpha} = \VBoth$.
\end{tabular}
\]
Now assume that $\mathbf{A}$ is a counter-model for $\Gamma \pEnt A$.
Then, it follows straightforwardly from its above-mentioned property 
that $\mathbf{A\sp{*}}$ is a counter-model for 
$\Embed{\Gamma \pEnt A}{}{}$.
From this, using the soundness of the proof system of $\FOCL(\TSigma)$, 
the if part follows immediately.
\qed
\end{proof}
From the property of the structure $\mathbf{A\sp{*}}$ of 
\smash{$\FOCL(\TSigma)$} referred to in the proof of 
Theorem~\ref{theorem-embedding}, it follows immediately that, 
as anticipated, for all atomic formula $A \in \SAtom{\Sigma}$:
\[
\renewcommand{\arraystretch}{1.25}
\begin{tabular}[t]{rl}
$\Term{A}{\mathbf{A\sp{*}}}{\alpha} = \VTrue$             & iff\,
$\Term{A}{\mathbf{A}}{\alpha} \in \set{\VTrue,\VBoth}$,
\\
$\Term{\denial{A}}{\mathbf{A\sp{*}}}{\alpha} = \VTrue$    & iff\,
$\Term{\Not A}{\mathbf{A}}{\alpha} \in \set{\VTrue,\VBoth}$,
\\
$\Term{A}{\mathbf{A\sp{*}}}{\alpha} = \VTrue$ & or\,
$\Term{\denial{A}}{\mathbf{A\sp{*}}}{\alpha} = \VTrue$.
\end{tabular}
\]
The following is a corollary of Theorems~\ref{theorem-sound-complete} 
and~\ref{theorem-embedding}.
\begin{corollary}
\label{corollary-embedding}
For all finite $\Gamma \subseteq \SForm{\Sigma}$ and 
$A \in \SForm{\Sigma}$:
\vspace*{-1.125ex} \par
\[
\renewcommand{\arraystretch}{1.125}
\begin{tabular}[t]{rl}
$\Gamma \pEnt A$ is valid &
 iff $\Embed{\Gamma \pEnt A}{}{}$ is classically valid;
\\
$\Gamma \pEnt A$ is satisfiable &
 iff $\Embed{\Gamma \pEnt A}{}{}$ is classically satisfiable.
\end{tabular}
\]
\end{corollary}

The translation of sequents extends to inference rules in the obvious 
way.
\begin{corollary}
\label{corollary-embedding-rules}
The translation of the inference rules of the presented proof system of 
$\foLPif(\Sigma)$ are derived inference rules of the proof system of  
$\FOCL(\TSigma)$ described in Section~\ref{subsect-FOCL}.
\end{corollary}

Seeing the translation of formulas of the form 
$\Not (A\sb1 \IImpl A\sb2)$, one might at first sight doubt whether the 
given translation provides an embedding of $\foLPif(\Sigma)$ into 
$\FOCL(\TSigma)$.
After all, $A_1 \IImpl A_2 \LEqv \Not A_1 \Or A_2$ does not hold for the 
logical equivalence relation of \foLPif.  
However, the fact that 
$\LCon \Not (A_1 \IImpl A_2) \BImpl \Not(\Not A_1 \Or A_2)$
holds for the logical consequence relation of \foLPif\ is sufficient for 
the given translation to provide an embedding of $\foLPif(\Sigma)$ into 
$\FOCL(\TSigma)$.

There may be alternatives to the given translation of formulas of 
$\foLPif(\Sigma)$ to formulas of $\FOCL(\TSigma)$.
Useful properties of the given translation are that:
\begin{itemize}
\item 
the signature is only extended with predicate symbols;
\item
in the translated formulas, the connective $\Not$ occurs only in 
subformulas of the form $\Not A$ where $A$ is an atomic formula.
\end{itemize}

\subsection{Decidability of Validity and Satisfiability for Fragments}

Fragments of classical logic of which it is known that validity or 
satisfiability of sequents is decidable are usually restricted to 
signatures without function symbols of positive arity and sometimes also 
to formulas in which the connective $\Not$ occurs only in subformulas of
the form $\Not A$ where $A$ is an atomic formula.
This means that determining whether membership of such a fragment is 
preserved by the translation given above is usually facilitated by the 
above-mentioned properties of the given translation.

For example, it is easy to see that membership of the following 
fragments, among others, is preserved by the given translation:
FO$^2$, the two-variable fragment~\cite{Mor75a,GKV97a};
GF, the guarded fragment~\cite{Gra99a}; 
TGF, the triguarded fragment~\cite{RS18a}; 
Maslov's class $\overline{\mathrm{K}}$~\cite{Mas68a,FKM24a};
BSR, the Bernays–Sch\"onfinkel–Ramsey fragment~\cite{BS28a,Ram87a};
OF, the ordered fragment~\cite{Her90a}; 
UNF, the unary-negation fragment~\cite{CS13a};
GNF, the guarded-negation fragment~\cite{BCS15a}; 
SF, the separated fragment~\cite{SVW16a};
FF, the forward fragment~\cite{Bed21a}; and
AF, the adjacent fragment~\cite{BKP23a}.

Knowing that validity or satisfiablity is decidable in $\FOCL(\TSigma)$ 
for a fragment and that membership of that fragment is preserved by the 
given translation is sufficient to conclude that it is also decidable in 
$\foLPif(\Sigma)$ for that fragment.
This means that the translation of the formulas of $\foLPif(\Sigma)$ to 
the formulas of $\FOCL(\TSigma)$ can be useful, among other things, to 
determine for a fragment for which validity or satisfiability is 
decidable in $\FOCL(\TSigma)$ whether it is also decidable in 
$\foLPif(\Sigma)$.
Moreover, when it comes to determining the complexity of the validity or
satisfiabiliy problem for a fragment in $\foLPif(\Sigma)$ or to 
designing an algorithm for it, this translation can be useful as well 
because it leads to only a polynomial increase in the length of formulas.

\section{Major Properties of \foLPif}
\label{sect-properties}

In this section, the major properties of \foLPif\ concerning its logical 
consequence relation and its logical equivalence relation are presented.

\subsection{The logical consequence relation of \foLPif}
\label{subsect-lconsequence}

Below, the properties of \foLPif\ concerning its logical consequence 
relation are presented that are generally considered to be desirable 
properties of a reasonable paraconsistent first-order logic.
The symbol $\clLCon$ is used to denote the logical consequence 
relation of \FOCL.

The following are properties of \foLPif\ concerning its logical 
consequence relation:
\begin{enumerate}
\item[(a)]
\foLPif\ is \emph{normal}, i.e.\
$\LCon$ is such that for all $\Gamma \subseteq \SForm{\Sigma}$,
$A_1, A_2, A_3 \in \SForm{\Sigma}$, and $A \in \SAtom{\Sigma}$:
\pagebreak[2]
\[
\renewcommand{\arraystretch}{1.125}
\begin{array}[t]{r@{\;}c@{\;}l@{\;\;}l}
A \not\LCon \Not A  &\mathrm{and} & \Not A \not\LCon A\;,
\\
\Gamma \LCon A_1 \CAnd A_2        & \mathrm{iff} &
\multicolumn{2}{l}
{\Gamma \LCon A_1 \;\mathrm{and}\;
 \Gamma \LCon A_2\;,}
\\
A_1 \COr A_2, \Gamma \LCon A_3    & \mathrm{iff} &
\multicolumn{2}{l}
{A_1, \Gamma \LCon A_3 \;\mathrm{and}\;
 A_2, \Gamma \LCon A_3\;,}
\\
\Gamma \LCon A_1 \IImpl A_2       & \mathrm{iff} & 
A_1, \Gamma \LCon A_2\;;
\\
\Gamma \LCon \Forall{x}{A}        & \mathrm{iff} & 
\multicolumn{2}{l}
{\Gamma \LCon A \hspace*{6em}
 \mathrm{provided}\; x \notin \ifree(\Gamma)\;,}
\\
\Gamma, \Exists{x}{A_1} \LCon A_2 & \mathrm{iff} & 
\multicolumn{2}{l}
{\Gamma, A_1 \LCon A_2 \hspace*{4em}
 \mathrm{provided}\; x \notin \ifree(\Gamma \union \set{A_2})\;;}
\end{array}
\]
\item[(b)]
\foLPif\ is $\Not$-\emph{contained in classical logic}, 
i.e.\ 
there exists a logic with the same logical constants, connectives, and 
quantifiers as \foLPif,
with the domain of truth values restricted to the classical truth values 
$\VTrue$ and $\VFalse$, and 
with a logical consequence relation $\LCon'$ such that:
\begin{itemize}
\item
${\LCon} \subseteq {\LCon'}$;
\item
$\Gamma \LCon' A$ iff,
for all structures $\mathbf{A}$,
for all assignments $\alpha$ in $\mathbf{A}$,
$\Term{A'}{\mathbf{A}}{\alpha} = \VFalse$ for some $A' \in \Gamma$ or
$\Term{A}{\mathbf{A}}{\alpha} = \VTrue$, where
the interpretation $\Term{\ph}{\ph}{\ph}$ is such that, 
for all structures $\mathbf{A}$,
for all assignments $\alpha$ in $\mathbf{A}$:
\[
\renewcommand{\arraystretch}{1.25}
\begin{array}[t]{c}
\Term{\Not A}{\mathbf{A}}{\alpha} =
 \left \{
 \begin{array}{l@{\;\;}l}
 \VTrue  & \mathrm{if}\; \Term{A}{\mathbf{A}}{\alpha} = \VFalse   \\
 \VFalse & \mathrm{if}\; \Term{A}{\mathbf{A}}{\alpha} = \VTrue\;, 
 \end{array}
 \right.
\end{array}
\]
where $a$ ranges over all truth values in $\set{\VTrue,\VFalse}$;
\end{itemize}
\item[(c)]
the propositional fragment of \foLPif\ is \emph{weakly maximal 
paraconsistent relative to classical logic}, i.e.\
for all $A \in \SProp{\Sigma}$ with $\nLCon A$ and $\clLCon A$, for the 
minimal consequence relation $\extLCon$ with
${\LCon} \subseteq {\extLCon}$ and $\extLCon A$, for all formulas 
$A' \in \SProp{\Sigma}$, $\extLCon A'$ iff $\clLCon A'$;
\item[(d)]
the propositional fragment of \foLPif\ is \emph{strongly maximal 
absolute paraconsistent}, i.e.\  
for all logics $\mathcal{L}$ with the same logical constants and 
connectives as \foLPif\ and with a consequence relation $\extLCon$ such 
that
$\set{\Gamma \LCon A \where
      \Gamma \union \set{A} \subseteq \SProp{\Sigma}} \subset
 \set{\Gamma \extLCon A \where
      \Gamma \union \set{A} \subseteq \SProp{\Sigma}}$,\,
$\mathcal{L}$ is not paraconsistent;
\item[(e)]
\foLPif\ \emph{enables internalization of consistency}, i.e.\ 
$A$ is consistent iff 
\mbox{${} \LCon (A \IImpl \False) \COr (\Not A \IImpl \False)$};
\item[(f)]
\foLPif\ \emph{enables internalization of logical equivalence}, i.e.\ 
$A_1 \LEqv A_2$ iff
\mbox{${} \LCon (A_1 \BIImpl A_2) \CAnd (\Not A_1 \BIImpl \Not A_2)$}.
\end{enumerate}

Properties~(a)--(c) indicate that \foLPif\ retains much of first-order
classical logic.
Properties~(a)--(c) and~(d) make the propositional fragment of \foLPif\ 
an ideal paraconsistent logic according to Definition~21 
in~\cite{AAZ11b}.
By property~(e), the propositional fragment of \foLPif\ is also a logic 
of formal inconsistency according to Definition~23 in~\cite{CCM07a}.

All three-valued paraconsistent propositional logics with the same 
logical constants, connectives, and quantifiers as \foLPif\ that have 
properties~(a) and~(b) have properties~(c)--(f) as well 
(cf.~\cite{Mid17a}).

From Theorem~4.42 in~\cite{AA15a}, it is known that there are exactly 
8192 different three-valued paraconsistent propositional logics with 
properties~(a) and (b).
From Corollary~4.74 in~\cite{AA15a}, it is known that the propositional 
fragment of \foLPif\ is the strongest three-valued paraconsistent 
propositional logic with property~(b) in the sense that for each 
three-valued paraconsistent propositional logic with property~(b) there
exists a logical consequence preserving translation of its formulas into 
formulas of the propositional fragment of \foLPif.

\subsection{The logical equivalence relation of \foLPif}
\label{subsect-lequivalence}

There are infinitely many different three-valued paraconsistent 
first-order logics with properties~(a) and (b).
This means that these properties, which concern the logical consequence 
relation of a logic, have no discriminating power.
The same holds for properties~(c)--(f) because each three-valued 
paraconsistent first-order logics with properties~(a) and~(b) has these 
properties as well.

Below, properties concerning the logical equivalence relation of a logic 
are used for discrimination.
It turns out that 13 classical laws of logical equivalence that also 
hold for the logical equivalence relation of \foLPif\ are sufficient to 
distinguish \foLPif\ completely from all other three-valued 
paraconsistent first-order logics with properties~(a) and~(b).

The laws in question are the identity, annihilation, idempotent, and 
commutative laws for conjunction and disjunction, the double negation 
law, two laws that uniquely characterize implication, and two laws that
concern universal and existential quantification.
%
%
\begin{theorem}
\label{theorem-soundness}
The logical equivalence relation of \foLPif\ satisfies laws (1)--(13) 
from Table~\ref{laws-lequiv}.
\begin{table}[!t]
\caption{Distinguishing laws of logical equivalence for \foLPif}
\label{laws-lequiv}
\begin{eqntbl}
\begin{eqncol}
(1)  & A \And \False \LEqv \False \\
(3)  & A \And \True \LEqv A \\
(5)  & A \And A \LEqv A \\
(7)  & A_1 \And A_2 \LEqv A_2 \And A_1 \\
(9)  & \Not \Not A \LEqv A \\
     & \\
(12) & \Forall{x}{(A_1 \And A_2)} \LEqv (\Forall{x}{A_1}) \And A_2 \\
     & \hfill \mathrm{if}\; x \notin \ifree(A_2) 
\end{eqncol}
\qquad\;
\begin{eqncol}
(2)  & A \Or \True \LEqv \True \\
(4)  & A \Or \False \LEqv A \\
(6)  & A \Or A \LEqv A \\
(8)  & A_1 \Or A_2 \LEqv A_2 \Or A_1 \\
(10) & \False \Impl A \LEqv \True \\
(11) & (A_1 \Or \Not A_1) \Impl A_2 \LEqv A_2 \\
(13) & \Exists{x}{(A_1 \Or A_2)} \LEqv (\Exists{x}{A_1}) \Or A_2 \\
     & \hfill \mathrm{if}\; x \notin \ifree(A_2) 
\end{eqncol}
\end{eqntbl}
\end{table}
\end{theorem}
\begin{proof}
For each of the laws~(1)--(13), with the exception of law~(11), 
satisfaction follows directly from the definition of the interpretation 
function for formulas given in Table~\ref{table-interpretation}.
For law~(11), we first have to establish that
$\Term{A \COr \Not A}{\mathbf{A}}{\alpha} \in \set{\VTrue,\VBoth}$.
\qed
\end{proof}
Moreover, among the infinitely many three-valued paraconsistent 
first-order logics with properties~(a) and~(b), \foLPif\ is the only 
one whose logical equivalence relation satisfies all laws given in 
Table~\ref{laws-lequiv}.\pagebreak[2]%
\footnote
{The paracomplete analogue of \foLPif\ is the only three-valued
 paracomplete first-order logic with properties~(a) and~(b) whose
 logical equivalence relation satisfies the laws from 
 Table~\ref{laws-lequiv}, with laws~(10) and ~(11) replaced by
 (10$'$)~$\True \Impl A \LEqv A$, and 
 (11$'$)~$(A_1 \And \Not A_1) \Impl A_2 \LEqv \True$~(cf.~\cite{Mid17a}).}
\begin{theorem}
\label{theorem-uniqueness}
There is exactly one three-valued paraconsistent first-order logic with 
properties~(a) and~(b) of which the logical equivalence relation 
satisfies laws (1)--(13) from Table~\ref{laws-lequiv}.
\end{theorem}
\begin{proof}
We know from Theorem~4.2 in~\cite{Mid17a} that for each of the logical
connectives there are laws among laws (1)--(11) that exclude all but one 
of its possible interpretations.
Moreover, given the remaining interpretations of $\And$ and $\Or$, it is 
not hard to see that laws~(12) and~(13) cannot hold if the interpretations 
$\forall$ and $\exists$ differ from their interpretations in \foLPif.
\qed
\end{proof}

It follows immediately from property~(a) that the logical equivalence 
relation of every three-valued paraconsistent first-order logics with 
properties~(a) and~(b) satisfies law~(1) from Table~\ref{laws-lequiv}.
It follows immediately from the proof of 
Theorem~\ref{theorem-uniqueness} that all proper subsets of laws 
(2)--(13) from Table~\ref{laws-lequiv} are insufficient to distinguish 
\foLPif\ completely from the other three-valued paraconsistent 
first-order logics with properties~(a) and~(b).

The next corollary also follow immediately from the proof of 
Theorem~\ref{theorem-uniqueness}.
\begin{corollary}
\label{corollary-exactly-one}
There are exactly 16 three-valued paraconsistent first-order logics 
with properties~(a) and~(b) of which the logical equivalence relation 
satisfies laws (1)--(9), (12), and~(13) from Table~\ref{laws-lequiv}.
\end{corollary}

It should be mentioned that the logical equivalence relation of \foLPif\ 
does not only satisfy the identity, annihilation, idempotent and 
commutative laws for conjunction and disjunction but also other basic 
classical laws for conjunction and disjunction, including the absorption, 
associative, distributive and de Morgan’s laws (cf.~\cite{Mid17a}).

\subsection{On the closeness of \foLPif\ to \FOCL}

Below, the different properties \foLPif\ related to closeness to \FOCL\
are briefly discussed.

\foLPif\ is a paraconsistent logic whose properties concerning its 
logical consequence relation include virtually all properties that have 
been proposed as desirable properties of such a logic.
Most properties concerned are related to closeness to~\FOCL.

If closeness to \FOCL\ is considered important, the above-mentioned 
properties concerning the logical equivalence relation concerning 
conjunction, disjunction, negation, universal quantification and 
existential should arguably also be taken as desirable properties of a 
paraconsistent logic.

Moreover, \foLPif\ has no connective or quantifier that is foreign to 
\FOCL\ and the inference rules of its natural deduction proof system are 
all known from \FOCL:
\begin{itemize}
\item
except for the inference rules concerning the negation connective, the 
inference rules are the ones found in all natural deduction proof 
systems for \FOCL;
\item
the inference rules concerning the negation connective are a rule that 
corresponds to the law of the excluded middle and rules that correspond 
to the de~Morgan's laws for all connectives and quantifiers;
\item
the rule corresponding to the law of the excluded middle is also found 
in natural deduction proof systems for \FOCL\ and the rules 
corresponding to the de~Morgan's laws are well-known derived rules of 
natural deduction proof systems for \FOCL.
\end{itemize}
This means that natural deduction reasoning in the setting of \foLPif\ 
differs from classical natural deduction reasoning only by slightly
different, but classically justifiable, reasoning about negations.

The translation that provides an embedding of \foLPif\ into \FOCL\ also 
shows the closeness of \foLPif\ to \FOCL.
Every formula of \foLPif, like every formula of \FOCL, has a negation 
normal form.
For a formula in negation normal form, the translation causes only minor 
changes. 
The translation consists solely of replacing each subformula of the form
$\Not A$, where $A$ is an atomic formula, by $\Not A \Or \denial{A}$.

\section{Concluding Remarks}
\label{sect-concl}

The paraconsistent first-order logic \foLPif\ has been presented.
A sequent-style natural deduction proof system has been given for this
logic. 
In addition to the model-theoretic justification of the proof system, a 
logical justification by means of an embedding into classical logic has 
been given.
This embedding also provides a classical-logic explanation of \foLPif.

In~\cite{JM93a}, an embedding of a paracomplete first-order logic into
classical first-order logic by means of a translation of its sequents 
has been given.
Because the approach followed in that paper is likely to work for all 
truth-functional finitely-valued logics, it was also followed in the 
first versions of the current paper. 
Removal of unnecessary complexity from the original translation has
resulted in the simpler translation presented in the current version of 
the paper.
This simpler translation turns out to be a generalization of the 
translation given for the propositional fragment of \CLuNs\ 
in~\cite{BCK99a}.

In~\cite{BM15b}, an application of the propositional fragment of 
\foLPif\ in the area of process algebra can be found.
That application concerns a process algebra in which propositions are 
used as the visible part of the state of processes. 
The composition of processes is dealt with in a way based on \foLPif.
This makes it possible, among other things, not to treat the composition 
of processes with contradictory visible states as an exception.
\foLPif\ is one of the four applicable three-valued logics that is 
normal and $\Not$-contained in classical logic (see 
Section~\ref{subsect-lconsequence}, properties~(a) and (b)).
\foLPif\ is even the only applicable one where, in addition, the law 
$\Not \Not A \LEqv A$ holds.

In~\cite{Mid22b}, an application of full \foLPif\ in the area of 
relational database theory can be found.
In that application, consistent query answering with respect to a 
possibly inconsistent database is considered from the perspective of
\foLPif.
This makes it possible, among other things, to define a consistent 
answer to a query with respect to a possibly inconsistent database 
without resort to database repairs.
The definitions and results are essentially the same as the 
corresponding definitions and results from the widely accepted classical 
logic based view of Reiter~\cite{Rei84a} if only consistent relational 
databases are considered.
This does not seem to be possible with any other three-valued logic that 
is normal and \mbox{$\Not$-contained} in classical logic.
 
The above-mentioned experiences with applications of \foLPif\ strengthen 
the impression that \foLPif\ is among the paraconsistent logics that 
deserve most attention.
However, the question arises whether a paraconsistent logic is really
needed to deal with contradictory sets of formulas.
The embedding of \foLPif\ into \FOCL\ given in this paper shows that it
can be dealt with in classical logic but in a much less convenient way.

In this paper, a sequent-style natural deduction proof system of 
\foLPif\ is presented.
In~\cite{Mid22b}, a sequent calculus proof system of \foLPif\ is 
presented.

\bibliographystyle{splncs04}
\bibliography{PCL}

\end{document}